\documentclass[twocolumn,superscriptaddress,prl,aps,preprintnumbers,nofootinbib,longbibliography]{revtex4-2}
\usepackage{amsmath,amssymb,bm,graphicx,color,gensymb,bbold,appendix}
\usepackage[colorlinks=true, citecolor={blue!80!black}, urlcolor={blue!50!black}, linkcolor = {blue!80!black}]{hyperref}
\usepackage[utf8x]{inputenc}

\usepackage{xcolor}

\newcommand{\CCoB}{Cs$_2$CoBr$_4$}
\newcommand{\CCoC}{Cs$_2$CoCl$_4$}
\newcommand{\CCuB}{Cs$_2$CuBr$_4$}
\newcommand{\CCuC}{Cs$_2$CuCl$_4$}

\newcommand{\BACOVO}{BaCo$_2$V$_2$O$_8$}
\newcommand{\SRCOVO}{SrCo$_2$V$_2$O$_8$}
\newcommand{\YAO}{YbAlO$_3$}

\newcommand{\ket}[1]{\left| #1 \right\rangle}

\newcommand{\hamilt}{{\hat{\mathcal{H}}}}

\usepackage{pdfpages} 
\usepackage{pgffor} 

\makeatletter
\AtBeginDocument{\let\LS@rot\@undefined}
\makeatother

\pdfximage{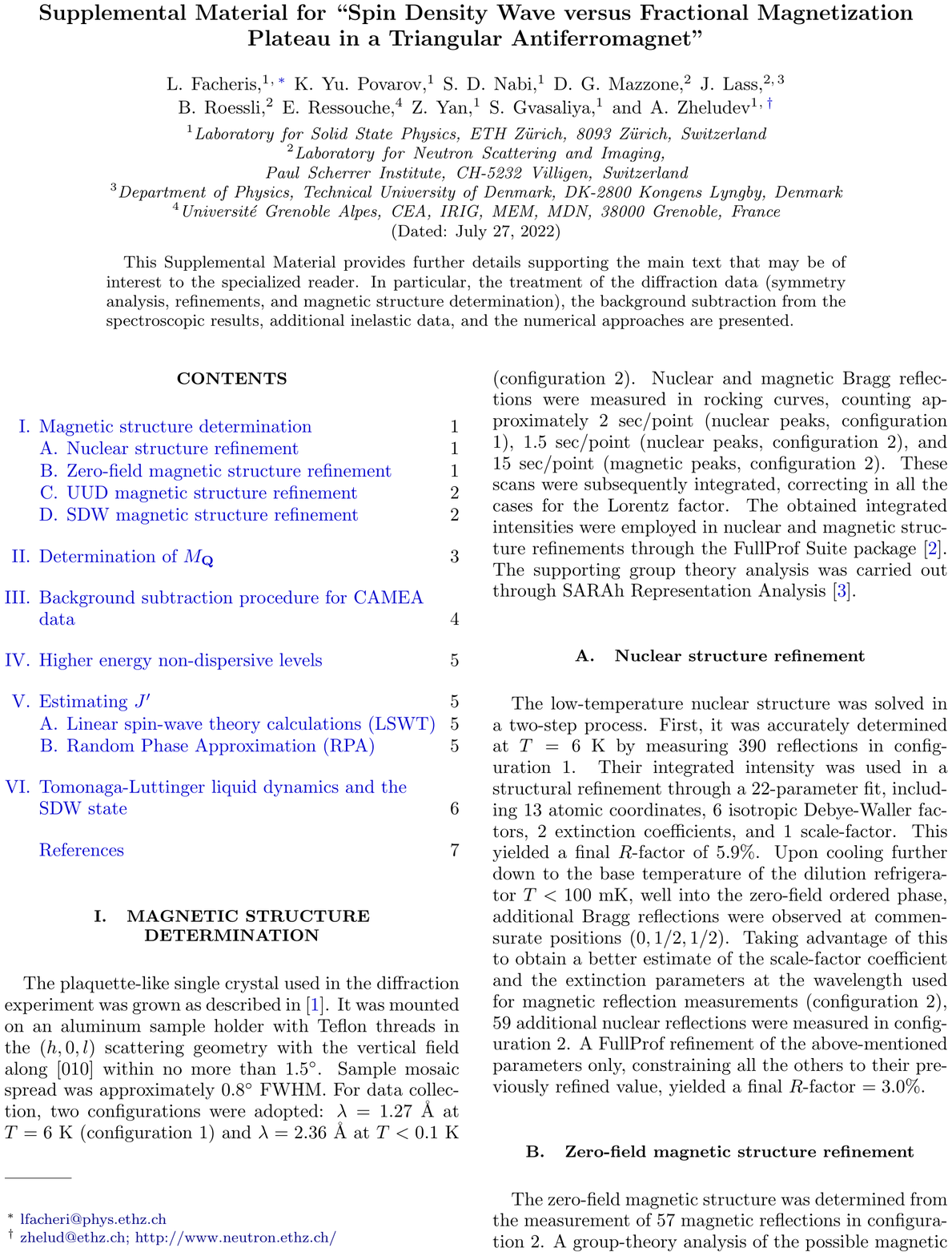}
\def\numbersupplementpages{\the\pdflastximagepages}

\begin{document}

\title{Spin density wave versus fractional magnetization plateau in a triangular antiferromagnet}
\author{L.~Facheris}
\email{lfacheri@phys.ethz.ch}
\address{Laboratory for Solid State Physics, ETH Z\"{u}rich, 8093 Z\"{u}rich, Switzerland}
\author{K.~Yu.~Povarov}
\address{Laboratory for Solid State Physics, ETH Z\"{u}rich, 8093 Z\"{u}rich, Switzerland}
\author{S.~D.~Nabi}
\address{Laboratory for Solid State Physics, ETH Z\"{u}rich, 8093 Z\"{u}rich, Switzerland}

\author{D.~G.~Mazzone}
\address{Laboratory for Neutron Scattering and Imaging, Paul Scherrer Institute, CH-5232 Villigen, Switzerland}
\author{J.~Lass}
\address{Laboratory for Neutron Scattering and Imaging, Paul Scherrer Institute, CH-5232 Villigen, Switzerland}
\address{Department of Physics, Technical University of Denmark, DK-2800 Kongens Lyngby, Denmark}
\author{B.~Roessli}
\address{Laboratory for Neutron Scattering and Imaging, Paul Scherrer Institute, CH-5232 Villigen, Switzerland}

\author{E.~Ressouche}
\address{Université Grenoble Alpes, CEA, IRIG, MEM, MDN, 38000 Grenoble, France}

\author{Z.~Yan}
\address{Laboratory for Solid State Physics, ETH Z\"{u}rich, 8093 Z\"{u}rich, Switzerland}
\author{S.~Gvasaliya}
\address{Laboratory for Solid State Physics, ETH Z\"{u}rich, 8093 Z\"{u}rich, Switzerland}
\author{A.~Zheludev}
\email{zhelud@ethz.ch; http://www.neutron.ethz.ch/}
\address{Laboratory for Solid State Physics, ETH Z\"{u}rich, 8093 Z\"{u}rich, Switzerland}

\begin{abstract}
We report an excellent realization of the highly non-classical incommensurate spin-density wave (SDW) state in the quantum frustrated antiferromagnetic insulator \CCoB. In contrast to the well-known Ising spin chain case, here the SDW is stabilized by virtue of competing planar in-chain anisotropies and frustrated interchain exchange. Adjacent to the SDW phase is a broad $m=1/3$ magnetization plateau that can be seen as a  commensurate locking of the SDW state into the up-up-down spin structure. This represents the first example of long-sought SDW-UUD transition in triangular-type quantum magnets.
\end{abstract}

\date{\today}
\maketitle

Of the various magnetically ordered phases of insulators, the spin-density wave (SDW) is perhaps the  least classical one~\cite{Giamarchi_2004_1Dbook,Starykh_RepPrPhys_2015_TriangularReview}. While easily envisioned in metals where the spin carriers are itinerant~\cite{Gruner_RPM_1994_metalSDW}, it cannot exist in classical models with localized spins of given magnitude at each site. Nevertheless, there are several purely quantum-mechanical routes to realizing SDW states in insulators. To a greater or lesser extent they are all based on the Tomonaga--Luttinger Spin Liquid (TLSL) properties of the $S=1/2$ quantum spin chain with antiferromagnetic (AF) exchange interactions $J$. In applied magnetic fields, a single chain develops incommensurate spin correlations in the longitudinal channel~\cite{Haldane_PRL_1980_TLSLconjecture,*BogolyubovIzerginKorepin_NucPhysB_1986_TLLexponents,*HikiharaFuruski_PRB_2004_XXZtll,OkunishiSuzuki_PRB_2007_XXZSDW}. In most cases though, the transverse commensurate correlations dominate, eventually resulting in transverse AF or helical long-range order in coupled chains. To create a SDW, one needs to somehow boost the longitudinal  correlations in each chain or to ensure that they are favored by inter-chain interactions.
The first approach, realized in materials like \BACOVO~\cite{KimuraMatsuda_PRL_2008_BACOVOsdw1,CanevetGrenier_PRB_2013_BACOVOsdw2} and \SRCOVO~\cite{ShenZaharko_NewJPhys_2019_SrCOVOdiffractionSDW}, is to simply endue the chains with Ising-type anisotropy. Another route, realized quite recently~\cite{Agrapidis_PRB_2019_YbAlO3modelSDW,FanYang_PRR_2020_IsingCoupledChains,*FanYu_ChinPhysB_2020_IsingCoupledChains}, is to impose Ising anisotropy on \emph{interchain} interactions $J'$. This is arguably the case of \YAO~\cite{WuNikitin_NatComm_2019_YbAlOSDW,NikitinNishimoto_NatComm_2021_YbAlOsolitons}. The third route to a SDW state exploits \emph{frustrated} zig-zag interchain bonds $J'$ in the so-called ``distorted triangular lattice'' geometry. Commensurate transverse TLSL correlations in each chain become completely decoupled at the Mean Field (MF) level. Incommensurate longitudinal ones are not, and are thus the ones to order in 3D. In this model theory predicts  a SDW phase in a very wide range of $J'/J$ ratios~\cite{Starykh_PRB_2010_Cs2CuCl4theory,ChenJuJiang_PRB_2013_DistortTriangquasi1D,StarykhBalents_PRB_2014_variousSDWexcitations}.  A very special feature of this mechanism is the SDW ``locking'' to a commensurate wavevector producing a $m=1/3$ up-up-down (UUD) magnetization plateau state. The latter persists even if $J=J'$, where no chains can be identified, and even in the fully isotropic case \cite{ChubukovGolosov_JPCM_1991_QuantumUUD}. It thus establishes an important link between 1D TLSL and 2D triangular lattice physics.
To date, this connection remains poorly understood experimentally, for lack of a suitable model system.

In the present Letter we demonstrate the existence of an incommensurate SDW state and its locking into a UUD phase in the triangular-lattice magnet \CCoB~\cite{PovarovFacheris_PRR_2020_CCoBplateaux}. Both phases are well-pronounced and occupy nearly a quarter of the phase diagram each. The mechanism behind this phenomenon is likely to be a blend of the three ``routes to SDW'' described above.

\begin{figure}
 \includegraphics[width=0.5\textwidth]{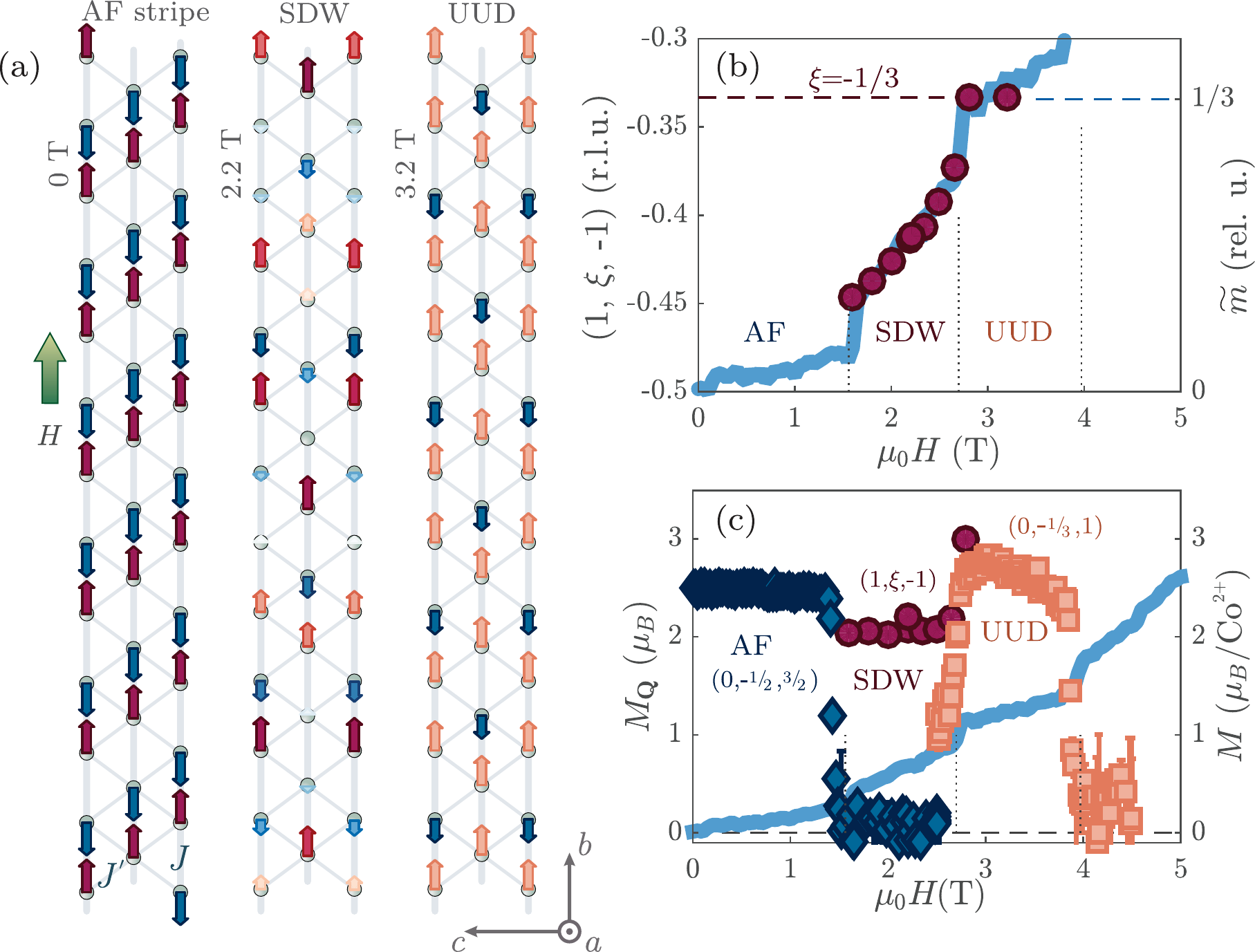}
\caption{Diffraction results for \CCoB, $\mathbf{H}\parallel\bm{b}$. (a) AF-stripe, SDW, and UUD phases shown in pseudospin representation. Color denotes the relative depth of on-site modulation. Exchange couplings $J$ and $J'$ are also indicated. (b) Points: incommensurate Bragg peak position of the SDW phase versus field. The blue line shows the ``pseudospin'' relative magnetization~\cite{PovarovFacheris_PRR_2020_CCoBplateaux}. (c) Magnetic order parameter (modulation amplitude) $M_\textbf{Q}$ vs field for AF (diamonds), SDW (circles), and UUD (squares) phases. The corresponding wave vectors of actual measurement are also indicated. The blue line is the magnetization curve $M(H)$ ~\cite{PovarovFacheris_PRR_2020_CCoBplateaux}. All measurements are performed at $T\lesssim0.1$~K.}\label{FIG:diffraction}
\end{figure}

\begin{figure}
 \includegraphics[width=0.5\textwidth]{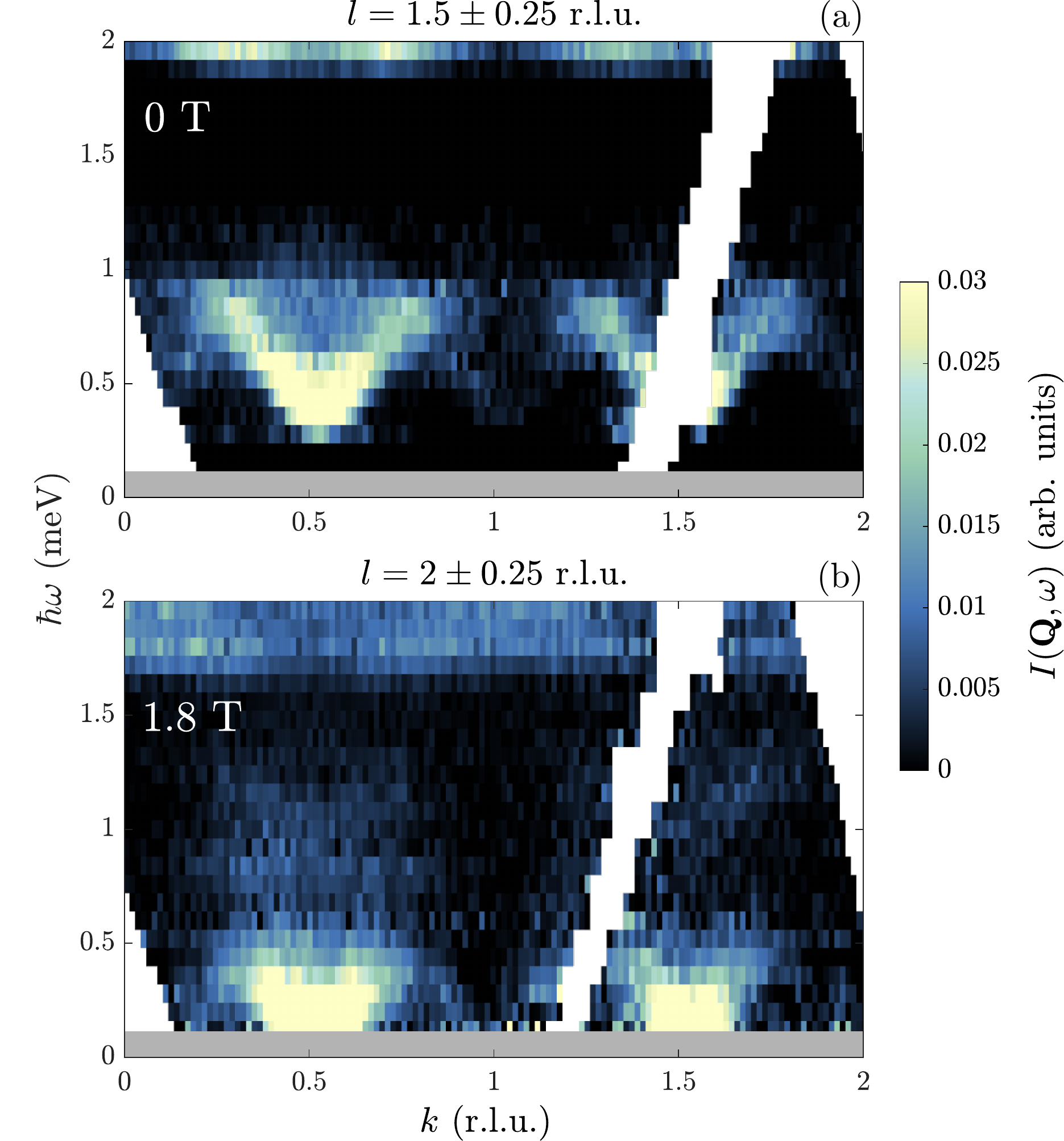}
 \caption{Overview of magnetic excitations in \CCoB\ at $T\lesssim0.1$~K in (a) zero field and (b) $1.8$~T. Color indicates the neutron scattering intensity proportional to the dynamic structure factor $\mathcal{S}(\mathbf{q},\omega)$. The data are integrated along $\mathbf{c}^\ast$ direction within the limits indicated. Background subtraction was performed as described in~\cite{SM}. Gray areas mask regions where the incoherent scattering dominates over the signal.}\label{FIG:CAMoverview}
\end{figure}

Our target material \CCoB\ crystallizes in the orthorhombic $P_{nma}$ structure, same as that of \CCuC~\cite{Coldea_PRL_2001_CCC2D,Tokiwa_PRB_2006_Cs2CuCl4phases}, \CCuB~\cite{OnoTanaka_JPSJ_2005_Cs2CuBr4plateuinvest}, and \CCoC~\cite{KenzelmannColdea_PRB_2002_CsCoCldiffraction,BreunigGarst_PRL_2013_CsCoCl_TF,*Breunig_PRB_2015_CsCoClPhD,LaurellScheie_PRL_2021_Cs2CoCl4entangl}. The magnetic $3d$ ions (four per unit cell) are arranged in triangular-patterned layers in the $bc$ plane. The two Cu-based compounds mentioned above are straightforward $S=1/2$ Heisenberg $J-J'$ model magnets~\cite{Starykh_PRB_2010_Cs2CuCl4theory}. The physics of  \CCoC\ and our material is more complicated. The Co$^{2+}$ magnetic ions sit in a low-symmetry distorted tetrahedral environment, thus their orbital momentum is quenched. Their magnetism is described in the language of $S=3/2$ spins dominated by crystal-field effects. The latter reduce the magnetism to \emph{pseudospin} $\tilde{S}=1/2$ degrees of freedom at low temperatures. In the pseudospin representation the interactions are strongly modified, yielding a nearly XY-type coupling. The effect is strongest for intra-chain $J$ interactions along the $\bm{b}$ axis. These anisotropic interactions are mostly a crystal field effect, in contrast to more symmetric situations where anisotropic interactions arise from direct spin-orbit entanglement~\cite{LinesCobalt, JackeliKhaliullin}. In \CCoC\ the zig-zag inter-chain interactions $J'$ are negligible, making the material a nearly-ideal XY chain~\cite{Gosuly_2016_PhDthesis,LaurellScheie_PRL_2021_Cs2CoCl4entangl}. A distinctive feature of \CCoB\ is that $J'$ is much stronger and comparable to $J$ ~\cite{PovarovFacheris_PRR_2020_CCoBplateaux}. This creates a novel type of frustration, in addition to the already-present geometric frustration of $J'$. The crystal structure dictates mutually perpendicular directions of the planar anisotropy in neighboring chains. The $\bm{b}$ direction is shared by both planes. In this way two easy-plane anisotropies conspire to effectively produce an easy-$\bm{b}$-axis anisotropy for the pseudospins.

Below $T_{N}=1.3$~K, as a function of magnetic field applied along the $\bm{b}$-direction, \CCoB\ goes through a sequence of five magnetic phases~\cite{PovarovFacheris_PRR_2020_CCoBplateaux}. The 1st and the 3rd phases in increasing fields are magnetization plateaux, with pseudospin magnetization $\tilde{m}\simeq0$ and $1/3$, correspondingly. In this study we use neutron diffraction to unambiguously identify these states as antiferromagnetic stripe phase (AF) with propagation vector $\mathbf{Q}=(0,~1/2,~1/2)$, and UUD phase with $\mathbf{Q}=(0,~1/3,~0)$. The intermediate 2nd magnetic state turns out to be a longitudinal incommensurate SDW with propagation vector $\mathbf{Q}=(0,~\xi,~0)$. 
The experiment was performed on the CEA-CRG D23 lifting-counter diffractometer at ILL (Grenoble, France). The $24.90(4)$~mg single crystal of \CCoB\ was mounted on the cold finger of a dilution refrigerator $T \lesssim 0.1$~K in a vertical $6$~T cryomagnet, with $\mathbf{H}\parallel \bm{b}$ and $ac$ being in the horizontal scattering plane. This setting allowed us to cover $|h| \leq 6$, $-1\leq k \leq 0$ and $|l| \leq 7$ r.l.u. portion of the $(h,k,l)$ reciprocal space with $\lambda=2.36$~\AA\ neutrons (PG002), in which the Bragg peaks of the types described above were collected in magnetic fields of $0$,~$2.2$, and $3.2$~T correspondingly. The symmetry-based group theory analysis and model refinement (using SARA$h$~\cite{Wills_PhysB_2000_SARAh} and FullProf Suite~\cite{RODRIGUEZCARVAJAL199355,SM}) suggest the collinear configurations shown in Fig.~\ref{FIG:diffraction}(a) to be the optimal solutions with $R$-factors $7.1$, $14.8$, and $11$\% for AF, SDW, and UUD correspondingly. The field dependencies of magnetic order parameters $M_\mathbf{Q}$ associated with each phase (\emph{modulation amplitudes} at the given propagation vectors $\mathbf{Q}$) are shown in Fig.~\ref{FIG:diffraction}(c). For AF and UUD phases, the Bragg peak intensity was measured by counting at fixed $\mathbf{Q}$ versus field. For the SDW phase, the peak positions and intensities were extracted from broad $k$-scans at 8 field values \cite{SM}. The AF order parameter disappears above the discontinuous AF-SDW transition at $1.5$~T in agreement with thermodynamics \cite{PovarovFacheris_PRR_2020_CCoBplateaux}. The apparent residual intensity at $(0, -1/2, 3/2)$ in Fig.~\ref{FIG:diffraction}(c) is due to a simple background model \cite{SM}. In contrast, the SDW-UUD transition seems to be a quintessential incommensurate-commensurate locking, with rather insignificant jump in the spin modulation depth. The residual intensity of $(0, -1/3, 1)$ below $2.8$~T is due to the poor resolution along the vertical $k$-direction. The $M_\mathbf{Q}$ extracted from the broad $k$-scan at $2.8$~T inside the UUD phase agrees well with the more precise UUD dataset.

The field dependence of the SDW propagation vector $(0,\xi,0)$ was extracted from the same $k$-scans mentioned above. The result is plotted in Fig.~\ref{FIG:diffraction}(b). Remarkably, the propagation vector closely follows the pseudospin relative  magnetization~\cite{PovarovFacheris_PRR_2020_CCoBplateaux}:
\begin{equation}\label{EQ:incZZ}
|\xi|=1/2-\tilde{m}/2.
\end{equation}
Such behavior is typical of longitudinal incommensurate correlations specific to $S=1/2$ chains. In the TLSL framework this incommensurability corresponds to a nesting vector that spans the Fermi sea of \emph{fractionalized spinon quasiparticles}~\cite{Giamarchi_2004_1Dbook}: $\xi=2k_F$. This picture may provide the basic idea for understanding the physics of \CCoB, but the analogy is difficult to extend beyond Eq.~\ref{EQ:incZZ}. The actual field dependence of magnetization shown in Fig.~\ref{FIG:diffraction}(b) is entirely different from that of a XXZ chain in longitudinal field~\cite{Giamarchi_2004_1Dbook,KimuraMatsuda_PRL_2008_BACOVOsdw1,CanevetGrenier_PRB_2013_BACOVOsdw2}.

Eq.~\ref{EQ:incZZ} also holds for coupled-chain models, the Heisenberg $J-J'$ model in particular~\cite{Starykh_PRB_2010_Cs2CuCl4theory,ChenJuJiang_PRB_2013_DistortTriangquasi1D,StarykhBalents_PRB_2014_variousSDWexcitations}. Similar types of field dependencies were previously also observed in the Ising chains~\cite{KimuraMatsuda_PRL_2008_BACOVOsdw1,CanevetGrenier_PRB_2013_BACOVOsdw2,ShenZaharko_NewJPhys_2019_SrCOVOdiffractionSDW}, or Ising-coupled Heisenberg chains~\cite{WuNikitin_NatComm_2019_YbAlOSDW,NikitinNishimoto_NatComm_2021_YbAlOsolitons}. A very narrow ``elliptical spiral'' phase with linear incommensuration-field dependence was also reported for the structurally similar Heisenberg $J-J'$ magnet \CCuC~\cite{Coldea_PRL_2001_CCC2D}. In all these cases, however, the UUD phase at $\xi=1/3$ is either absent altogether (Ising chains), or is very narrow compared to the SDW state (\YAO). In the latter case, theory does not predict any plateau for the corresponding Ising-coupled Heisenberg chain model~\cite{Agrapidis_PRB_2019_YbAlO3modelSDW,FanYang_PRR_2020_IsingCoupledChains,*FanYu_ChinPhysB_2020_IsingCoupledChains}. It has been proposed that the commensurate-incommensurate locking in \YAO\  might be the consequence of e.g. additional small interactions~\cite{NikitinNishimoto_NatComm_2021_YbAlOsolitons}.
The coexistence of a magnetization-scaled SDW and well-defined plateau is unique for  \CCoB.

To get more insight into the mechanism of the AF-SDW-UUD sequence of transitions, we have measured the magnetic excitation spectra at zero field (AF) and at $1.8$~T (SDW). The experiment was performed on the new CAMEA spectrometer at PSI (Switzerland)~\cite{MarkoGroitl_RevSciInstr_2018_CAMEA,*LassJakobsen_SoftX_2020_Mjolnir}. The $m=1.16$~g crystal of \CCoB\ was mounted on the cold finger of a dilution refrigerator with $bc$ in the scattering plane. A $1.8$~T horizontal magnet was used, with the direction of the field set along $\bm{b}$. The unique combination of the multiplexing capabilities of CAMEA, ``continuous angle'' data acquisition mode, and the open geometry of this horizontal magnet allowed us to obtain a detailed neutron scattering intensity dataset vs $(k,l,\hbar\omega)$ in both AF and SDW states.  Two measurement series were performed at each field, with $E_\mathrm{i}=5.1$ and $E_\mathrm{i}=3.6$~meV, for higher coverage and higher resolution correspondingly ($\sim 0.16$~meV FWHM). The projections from the cumulative datasets for AF ($\mu_0 H=0$) and SDW ($\mu_0 H=1.8$~T) phases are shown in Fig.~\ref{FIG:CAMoverview}.
In zero field the spectrum is gapped ($\Delta\simeq0.35$~meV) and mostly dispersive along the $\bm{b}$ direction, with the bandwidth approaching $0.7$~meV. In addition to the pronounced magnon-like excitation at low energy, a continuum with a sharp upper boundary is clearly visible. At higher energies, $E_{\mathrm{CF}}=2.1(1)$~meV, we observe a non-dispersive level (see~\cite{SM} for extra data), that can be understood as $\ket{1/2}\rightarrow\ket{3/2}$ transition of cobalt $S=3/2$ at energy $2D$. This gives $D=12.2(6)$~K, in good agreement with the susceptibility-based estimate of $14(1)$~K reported earlier~\cite{PovarovFacheris_PRR_2020_CCoBplateaux}.

\begin{figure}
 \includegraphics[width=0.5\textwidth]{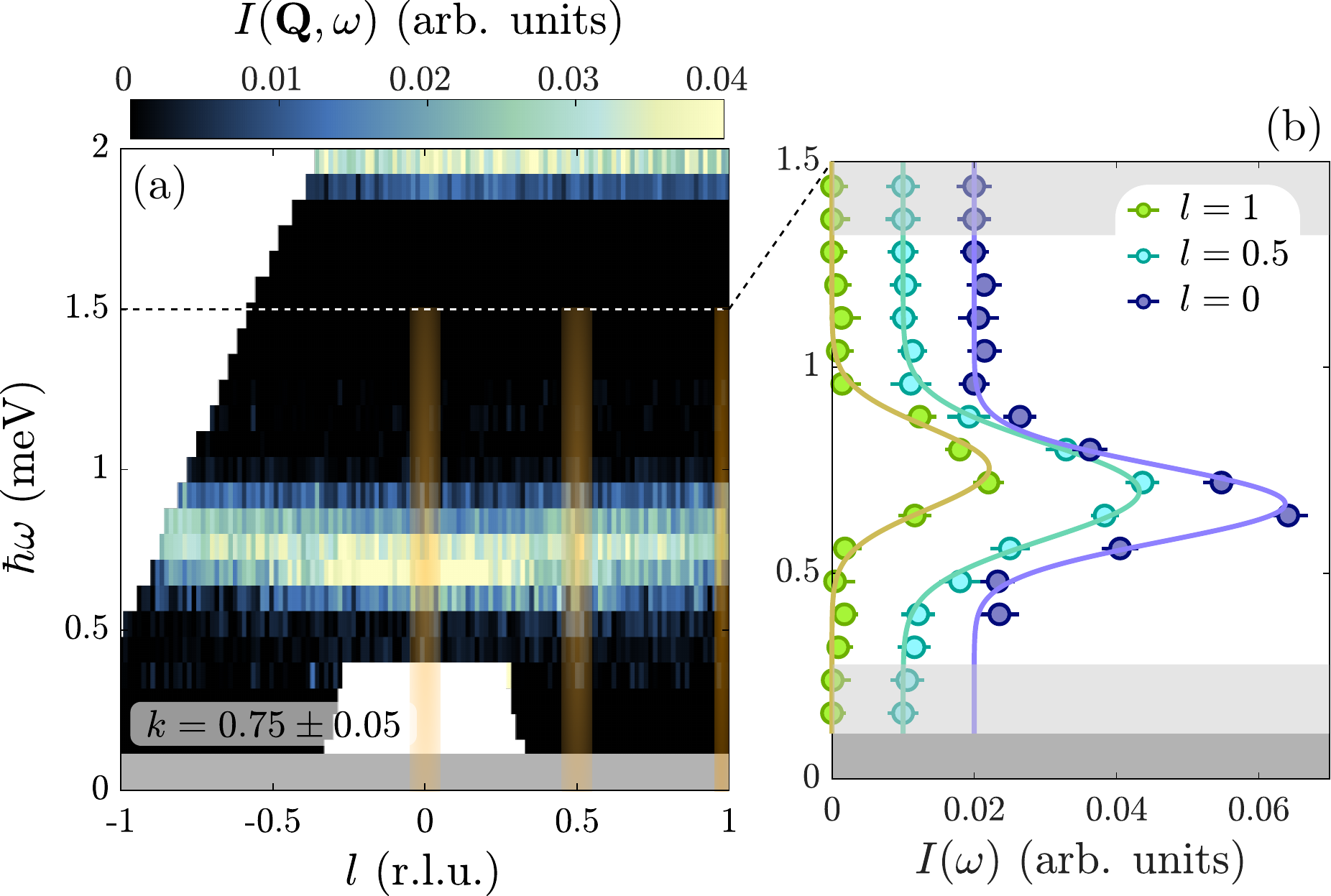}
 \caption{(a) Dispersion along the $\mathbf{c}^\ast$ direction measured in \CCoB~at $T\lesssim0.1$~K in zero field. The particular $k$ value and integration range are indicated in the plot. (b) Intensity vs. energy transfer cuts at different momenta, integrated in $0.05$ r.l.u. along $l$ (see orange stripes in (a)). Dark gray areas mask the incoherent scattering, while light gray bands hide points excluded from the fit.}\label{FIG:Lcut}
\end{figure}

The strength of $J'$ interactions is key to understand the physics of \CCoB.  As Fig.~\ref{FIG:Lcut} shows, the bandwidth along the $\mathbf{c}$ axis is only about $0.1$~meV.
This however, is not a sign of an insignificant $J'$ , but rather of geometric frustration in the zig-zag inter-chain coupling and of its predominant Ising nature.
For a crude estimate we can rely on a simple spin wave theory (SWT) calculation (using the SpinW package~\cite{TothLake_JPCM_2015_SpinW,SM}).
The starting point is the Hamiltonian
\begin{widetext}
\begin{equation}
\label{EQ:Hamilt}
\hamilt=\sum_{i,j}\sum_{\alpha=x,y,z}J^{\alpha\alpha}_{e}\hat{\tilde{S}}_{i,2j}^{\alpha}\hat{\tilde{S}}_{i+1,2j}^{\alpha}+J^{\alpha\alpha}_{o}\hat{\tilde{S}}_{i,2j+1}^{\alpha}\hat{\tilde{S}}_{i+1,2j+1}^{\alpha}+J^{\prime \alpha\alpha}\hat{\tilde{S}}_{i,2j}^{\alpha}[\hat{\tilde{S}}_{i,2j+1}^{\alpha}+\hat{\tilde{S}}_{i,2j-1}^{\alpha}],
\end{equation}
\end{widetext}
with $j$ and $i$ enumerating the chains and sites within, and the diagonal exchange tensors being $J^{\alpha\alpha}_{e,o}=J(1,1+\delta,1-\Delta_{\mathrm{XY}})$ and $J(1-\Delta_{\mathrm{XY}},1+\delta,1)$ for even and odd chains; $J^{\prime \alpha\alpha}=J'(1-\delta',1,1-\delta')$ for zig-zag bonds. Parameters $\Delta_\mathrm{XY}=0.75$ and $\delta'=0.5$ are fixed from the basic pseudospin representation arguments~\cite{BreunigGarst_PRL_2013_CsCoCl_TF,PovarovFacheris_PRR_2020_CCoBplateaux}. This is just a minor extension of the model presented earlier~\cite{PovarovFacheris_PRR_2020_CCoBplateaux}, which corresponds to $\delta=0$. As discussed in the Supplement, a good description of the main sharp spectral features is obtained with $J\simeq 0.8$~meV, $J'\simeq 0.35$~meV, and tiny additional Ising-type anisotropy $\delta=0.1$ present on the main $J$-bond. Since the semiclassical linear spin wave theory is ill-defined for systems involving strong quantum fluctuations like \CCoB, these results are a good approximation only of the low-energy spectrum \cite{SM,ColdeaTennant_fractionalizationCs2CuCl4,Tennant_KCuF3}. For this reason, we adopted an independent approach to estimate the ratio of $J'/J$: By using the Ising spin chain as the starting point, and then the Random Phase Approximation, the bandwidth along and transverse to the chain direction can be obtained~\cite{SM,JensenMackintosh_1991_RareEarthBook,KohnoStarykh_NatPhys_2007_Triplons}. This approach also points to a significant $J'/J\gtrsim0.4$ exchange ratio (see~\cite{SM}). We conclude that in \CCoB\ {\em the inter-chain coupling is almost half as strong as the in-chain one}. This is consistent with the observed spectrum being entirely different from that in weakly coupled Ising spin chains with a pronounced ``Zeeman ladder'' of spinonic bound states~\cite{Shiba_PTP_1980_ZeemanLadder}, such as in  \BACOVO~\cite{GrenierPetit_PRL_2015_BACOVOzeemanladders} .

\begin{figure}
 \includegraphics[width=0.5\textwidth]{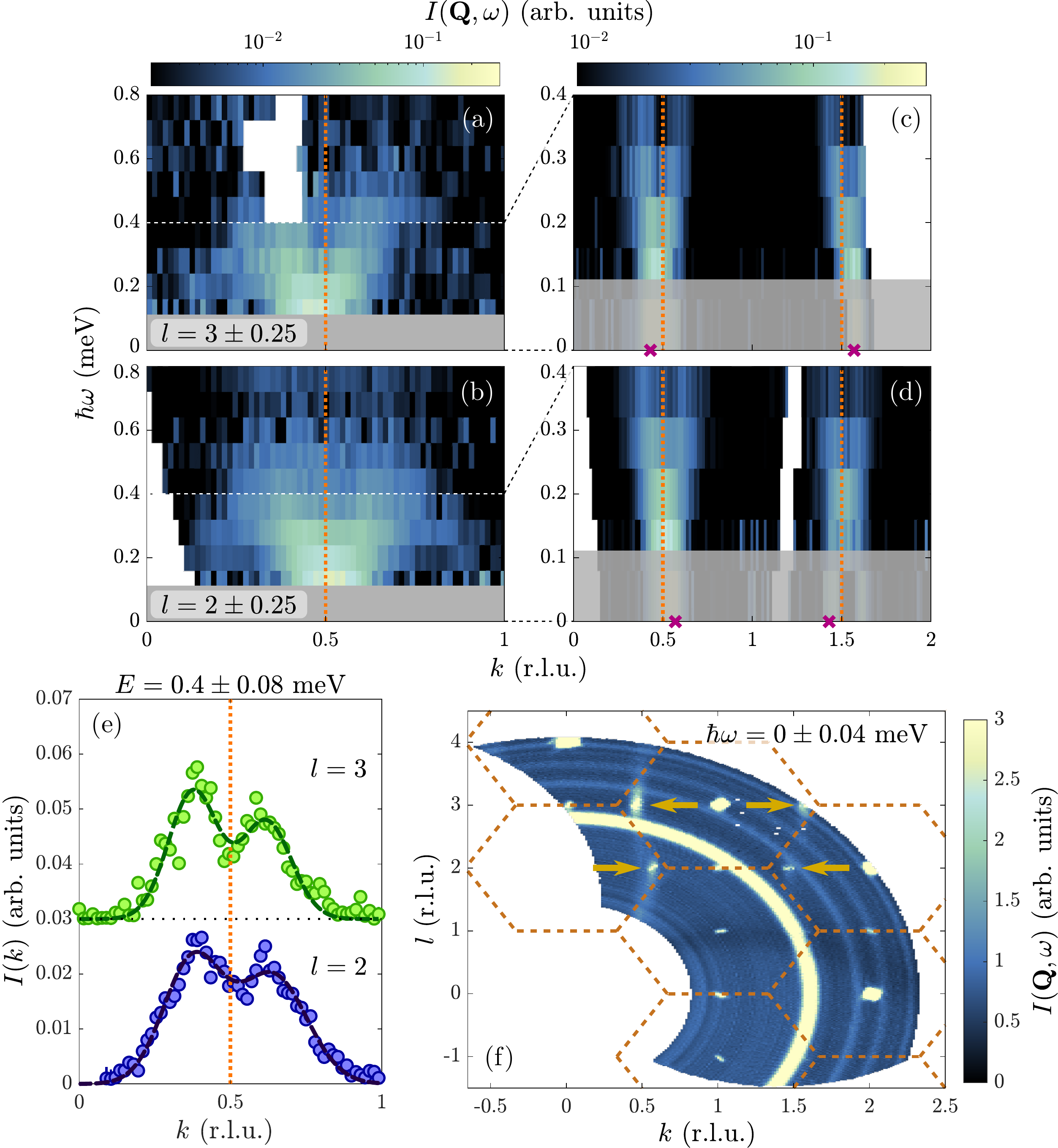}
 \caption{(a)-(b) Low-energy dynamics in the SDW phase of \CCoB~measured at $T\lesssim0.1$~K and $\mu_0H=1.8$~T. The panels show overviews of the ``continuum'' part at even and odd $l$ with (c)-(d) zoom into the lowest accessible energies correspondingly. Magenta crosses mark the positions of the magnetic Bragg peaks, orange dashed lines show the $k=1/2$ zone-centers. The scattering in grayed areas is dominated by the incoherent line. All data are background-substracted. (e) Intensity-momentum cut at two $l$-values. The energy integration range is shown. Dashed lines are double-gaussian fit. The $l =3$ curve is shifted upwards for visibility. (f) $0kl$-plane at $\hbar\omega = 0$~meV. Arrows indicate the positions of SDW Bragg peaks. Orange dashed lines mark the effective ``triangular'' Brillouin zones. Here the background is not subtracted.}\label{FIG:SDWstag}
\end{figure}

We turn to the dynamics of the SDW phase. The corresponding data set from CAMEA is actually quite unique, given the horizontal-field scattering geometry, the low energy scales and the wide reciprocal-space coverage. The data collected at $\mu_0H=1.8$~T are visualized in Fig.~\ref{FIG:CAMoverview}(b) and Fig.~\ref{FIG:SDWstag}. Compared to zero field, the spectrum measured in the SDW phase is much broader and clearly gapless. The latter agrees with the conclusions drawn from thermodynamic  measurements~\cite{PovarovFacheris_PRR_2020_CCoBplateaux}. As could be expected, the excitation energy goes to zero at the incommensurate $(0,\xi,0)$ positions of the SDW where the corresponding Bragg peaks are located. However, there is one crucial difference with the single-chain TLSL spectrum of longitudinal excitations. In the latter, the spectrum is symmetric with respect to the $k=1/2$ point. In \CCoB~it is not. The sign of $\xi$ in Eq.~\ref{EQ:incZZ} describing the Bragg peak and soft mode location corresponds to the parity of $l$ for odd $k\sim1/2$, but is reversed for $k\sim3/2$. This ``staggered'' pattern of soft modes is once again dictated by the quasi-triangular, rather than chain-like, nature of the underlying ionic lattice. Fig.~\ref{FIG:SDWstag}(f) shows this clearly. The low-energy intensity is condensed around the side edges of the hexagonal zones. The low-energy excitations with linear dispersion emanate from the corresponding Bragg peaks. Their shape at higher energy transfer appears symmetric around $k = 1/2$, as Fig.~\ref{FIG:SDWstag}(e) shows. The high energy part of the spectrum seems to be rather diffuse. We cannot identify any sharp modes above $0.4$~meV. This may be due to the splitting and smearing of the zero-field spectrum by non-commuting magnetic fields, although the scenario with spinon complex continua, characteristic of longitudinal fluctuations in the TLSL phase~\cite{Giamarchi_2004_1Dbook}, cannot be ruled out either. Moreover, the non-dispersive level has now shifted to $E'_{\mathrm{CF}}\simeq1.7(1)$~meV, in agreement with the expectations for easy-plane Co$^{2+}$ ions in transverse field with the anisotropy constant $D$ given above. To conclude, spectral properties of the SDW state are heavily affected by the 2D $J'$ exchange, but only at quite low energies. The high energy part seems to be a result of the interplay of spin-$1/2$ chain-like physics and Co$^{2+}$ crystal-field effects. Similar arguments apply to the high-field D and E phases \cite{PovarovFacheris_PRR_2020_CCoBplateaux}. As said, non-commuting fields admix $|\pm3/2\rangle$ states to the ground state, making the presudospin-$1/2$ model inadequate at finite field. Their gapless nature \cite{PovarovFacheris_PRR_2020_CCoBplateaux} hints to complex structure, among which are incommensurate planar states or spin-density waves \cite{Starykh_RepPrPhys_2015_TriangularReview}.

In conclusion, for the first time we have observed the generation of incommensurate longitudinal SDW phase and its locking into UUD magnetization plateau state, driven by quasi-2D correlations in a $J-J'$ distorted triangular lattice magnet. While there may exist some easy-axis anisotropy in both the intra- and interchain exchange, the frustration of zig-zag bonds appears to be the primary mechanism defining the phase diagram of magnetized \CCoB. This material appears to be an ideal platform for exploring the interplay of anisotropy and frustration. This exotic frustrated physics, interpolating between 1D and 2D worlds, calls for intense future experiments and in-depth theoretical analysis.

\acknowledgements

We acknowledge the beam time allocation at PSI and ILL (TASP id: 20200145, CAMEA id: 20211018, D23 id: 5-41-1086). J. Lass was supported by the Danish National Committee for Research Infrastructure through DanScatt. We thank Dr. Denis Golosov (Bar Ilan University) for insightful discussions.

\bibliography{SDW_UUD_Cs2CoBr4.bbl}

    

\section*{Supplementary material}

# Supplemental Material for "Spin Density Wave versus Fractional Magnetization Plateau in a Triangular Antiferromagnet"

L. Facheris,[1, *] K. Yu. Povarov,[1] S. D. Nabi,[1] D. G. Mazzone,[2] J. Lass,[2, 3]
B. Roessli,[2] E. Ressouche,[4] Z. Yan,[1] S. Gvasaliya,[1] and A. Zheludev[1, †]

[1]*Laboratory for Solid State Physics, ETH Zürich, 8093 Zürich, Switzerland*
[2]*Laboratory for Neutron Scattering and Imaging,
Paul Scherrer Institute, CH-5232 Villigen, Switzerland*
[3]*Department of Physics, Technical University of Denmark, DK-2800 Kongens Lyngby, Denmark*
[4]*Université Grenoble Alpes, CEA, IRIG, MEM, MDN, 38000 Grenoble, France*
(Dated: July 27, 2022)

This Supplemental Material provides further details supporting the main text that may be of interest to the specialized reader. In particular, the treatment of the diffraction data (symmetry analysis, refinements, and magnetic structure determination), the background subtraction from the spectroscopic results, additional inelastic data, and the numerical approaches are presented.

## CONTENTS

## I. MAGNETIC STRUCTURE DETERMINATION

The plaquette-like single crystal used in the diffraction experiment was grown as described in [1]. It was mounted on an aluminum sample holder with Teflon threads in the $(h, 0, l)$ scattering geometry with the vertical field along [010] within no more than 1.5°. Sample mosaic spread was approximately 0.8° FWHM. For data collection, two configurations were adopted: $\lambda = 1.27$ Å at $T = 6$ K (configuration 1) and $\lambda = 2.36$ Å at $T < 0.1$ K (configuration 2). Nuclear and magnetic Bragg reflections were measured in rocking curves, counting approximately 2 sec/point (nuclear peaks, configuration 1), 1.5 sec/point (nuclear peaks, configuration 2), and 15 sec/point (magnetic peaks, configuration 2). These scans were subsequently integrated, correcting in all the cases for the Lorentz factor. The obtained integrated intensities were employed in nuclear and magnetic structure refinements through the FullProf Suite package [2]. The supporting group theory analysis was carried out through SARAh Representation Analysis [3].

### A. Nuclear structure refinement

The low-temperature nuclear structure was solved in a two-step process. First, it was accurately determined at $T = 6$ K by measuring 390 reflections in configuration 1. Their integrated intensity was used in a structural refinement through a 22-parameter fit, including 13 atomic coordinates, 6 isotropic Debye-Waller factors, 2 extinction coefficients, and 1 scale-factor. This yielded a final $R$-factor of 5.9%. Upon cooling further down to the base temperature of the dilution refrigerator $T < 100$ mK, well into the zero-field ordered phase, additional Bragg reflections were observed at commensurate positions $(0, 1/2, 1/2)$. Taking advantage of this to obtain a better estimate of the scale-factor coefficient and the extinction parameters at the wavelength used for magnetic reflection measurements (configuration 2), 59 additional nuclear reflections were measured in configuration 2. A FullProf refinement of the above-mentioned parameters only, constraining all the others to their previously refined value, yielded a final $R$-factor = 3.0%.

### B. Zero-field magnetic structure refinement

The zero-field magnetic structure was determined from the measurement of 57 magnetic reflections in configuration 2. A group-theory analysis of the possible magnetic

---
[*] lfacheri@phys.ethz.ch
[†] zhelud@ethz.ch; http://www.neutron.ethz.ch/

structures compatible with the crystallographic symmetry and the propagation vector restricted the choice to two irreducible representations only, each containing 6 basis vectors. As in sister material $Cs_2CoCl_4$, which shares the same space group symmetry and zero-field propagation vector [4], only one was found to reproduce the data, whose wave-vectors are listed in Table I, according to the following convention

$$Co_1 = (0.262(1), 0.75, 0.580(1))$$
$$Co_2 = (0.762(1), 0.75, 0.920(1))$$
$$Co_3 = (0.738(1), 0.25, 0.420(1))$$
$$Co_4 = (0.238(1), 0.25, 0.080(1))$$

Assuming the spins to be confined to the triangular $bc$ plane and aligned along the crystallographic $b$ axis, the basis vectors in Table I would point to a structure of antiferromagnetic chains along $b$ with particular pattern between the chains (Fig. 1(a) in the main text). It must be noticed, however, that the irreducible representation under consideration can occur in several domains. To properly account for them all, each of the generators of the little group was applied separately to a given spin configuration:

$$GEN_1 = -x, -y, -z$$
$$GEN_2 = -x + 1/2, -y, z + 1/2$$
$$GEN_3 = -x, y + 1/2, -z$$

$GEN_1$ produces a structure where the $b$-component of *all* the spins in the unit cell is reversed with respect to the given configuration. This structure has thus the same structure factor as the original one, therefore it does not constitute an inequivalent domain for unpolarized neutrons. The action of $GEN_2$ and $GEN_3$ on the original configuration reverses by 180° the direction of the spins on $Co_1$ and $Co_2$, leaving $Co_3$ and $Co_4$ unaffected. This produces a configuration with a different structure factor than the original one, and therefore has to be considered as a possible domain.

Eventually, a magnetic refinement with 2 fitting parameters only (same ordered moments on all the sites with component only along $b$, and the relative population of domains) could account for the observed Bragg intensity with a R-factor = 7.1%. The result of the fit is shown in Supp. Fig. 1. The fitted parameters yield an ordered moment $m_0 = 2.44(2)$ $\mu_B$ with a relative contribution 0.45(1) of domains generated by $GEN_2$ and $GEN_3$.

Other possible solutions were considered. In particular, allowing the moments to tilt away from the $b$ direction in the $bc$ plane produced a slightly better solution (R-factor = 6.9%), but requires 1 fitting parameter more. Alternatively, allowing the spins out of the triangular plane yielded an out-of-plane component of the moments compatible with zero within the corresponding errorbar.

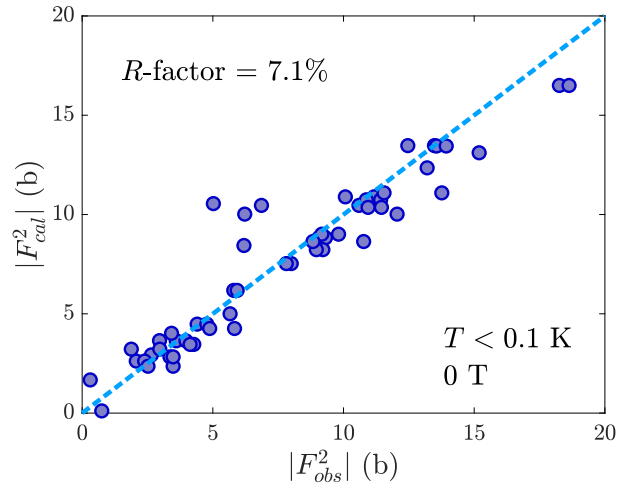

SUPP. FIG. 1. Refinement of the observed integrated intensities for magnetic Bragg peaks in zero-field against the model proposed in the text. A final R-factor of 7.1% was obtained.

### C. UUD magnetic structure refinement

At 3.2 T magnetic reflections were observed at positions indexed by the ordering vector $(0, 1/3, 0)$. A total of 71 was measured. From group-theory analysis, 4 possible irreducible representations with 3 basis vectors each were found. From magnetization measurements [1], the structure is expected to be compatible with a magnetization plateau state. For two of the irreps, no such structure is contemplated. Of the remaining two, one realizes the "up-up-down" state only for one of the triangular planes of the unit cell, with the other plane being "down-down-up". Therefore, the most promising candidate for the C phase is the remaining irreducible representation, whose basis vectors are listed in Table II. In this case, there are only 2 generators for the little group, which did not produce any inequivalent domain.

Assuming the structure to be collinear, with moments pointing only along the field, we can describe the structure with basis vector $\Psi_2$. A refinement through FullProf, with one fit parameter only (the ordered moment) and the magnetic phases on $Co_3$ and $Co_4$ constrained as in Table II quickly converged, with a final R-factor = 11% (Supp. Fig. 2). The obtained structure is of the UUD type, with an ordered moment of $m_0 = 2.67(2)$ $\mu_B$.

### D. SDW magnetic structure refinement

For the incommensurate phase, the allowed irreducible representations are similar to those in the UUD phase. Taking advantage of the locking transition, the realized irreducible representation is the same as in the UUD phase, with basis vectors as in Table II, with $\varphi = 73.8°$. Here, 71 Bragg reflections at 2.2 T were collected. As-

|     | $\Psi_1$ | $\Psi_2$ | $\Psi_3$ | $\Psi_4$ | $\Psi_5$ | $\Psi_6$ |
| --- | --- | --- | --- | --- | --- | --- |
| $Co_1$ | (1, 0, 0) | (0, 1, 0) | (0, 0, 1) | (1, 0, 0) | (0, −1, 0) | (0, 0, 1) |
| $Co_2$ | (1, 0, 0) | (0, −1, 0) | (0, 0, −1) | (1, 0, 0) | (0, 1, 0) | (0, 0, −1) |
| $Co_3$ | (1, 0, 0) | (0, 1, 0) | (0, 0, 1) | (−1, 0, 0) | (0, 1, 0) | (0, 0, −1) |
| $Co_4$ | (1, 0, 0) | (0, −1, 0) | (0, 0, −1) | (−1, 0, 0) | (0, −1, 0) | (0, 0, 1) |

TABLE I. Basis vector associated with the observed irreducible representation for each of the magnetic atoms in the crystallographic unit cell.

|     | $\Psi_1$ | $\Psi_2$ | $\Psi_3$ |
| --- | --- | --- | --- |
| $Co_1$ | (1, 0, 0) | (0, 1, 0) | (0, 0, 1) |
| $Co_2$ | (1, 0, 0) | (0, 1, 0) | (0, 0, −1) |
| $Co_3$ | $(e^{i\varphi}, 0, 0)$ | $(0, -e^{i\varphi}, 0)$ | $(0, 0, e^{i\varphi})$ |
| $Co_4$ | $(e^{i\varphi}, 0, 0)$ | $(0, -e^{i\varphi}, 0)$ | $(0, 0, -e^{i\varphi})$ |

TABLE II. Basis vector associated with the irreducible representation realized at 3.2 T, with $\varphi = 60°$.

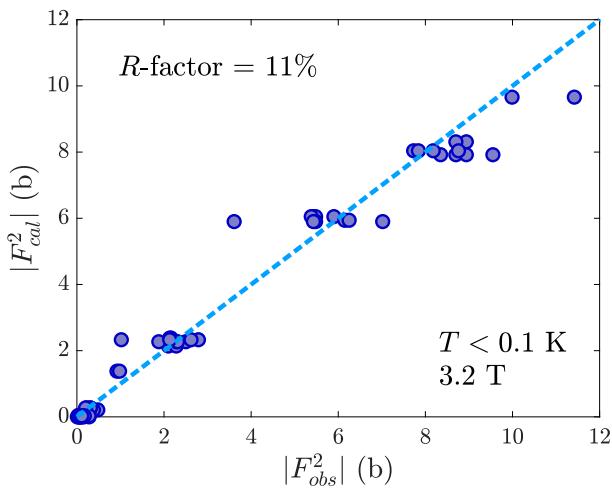

SUPP. FIG. 2. Refinement of the observed integrated intensities for magnetic Bragg peaks at 3.2 T. A final $R$-factor of 11% was obtained.

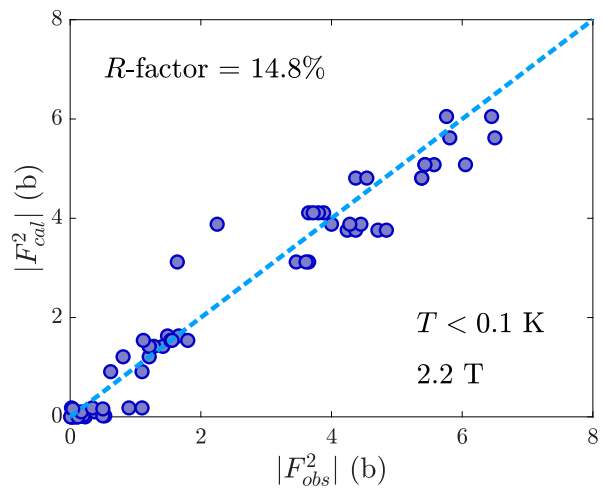

SUPP. FIG. 3. Refinement of the observed integrated intensities for magnetic Bragg peaks at 2.2 T. A final $R$-factor of 14.8% was obtained.

suming a purely collinear structure described again by $\Psi_2$, we can fix the magnetic phases to the ones prescribed by the propagation vector. A single-parameter FullProf refinement produced a longitudinal spin-density wave state with an $R$-factor = 14.8% (Supp. Fig. 3). In this case the ordered moment was found to be $m_0 = 2.14(3)$ $\mu_B$.

We consider another possible model based on a helical structure with the spins rotating in the triangular $bc$ plane according to $\Psi_2$ and $\Psi_3$ (2 refinement parameters). This yielded a similar solution in terms of $R$-factor = 14.8%, but the imaginary $c$-component of the moments was found to be compatible with 0.

## II. DETERMINATION OF $M_\mathbf{Q}$

This section describes how the data shown in Fig. 1(c) of the main text have been measured and how the corresponding $M_\mathbf{Q}$ has been extracted.

For the AF and UUD commensurate phases, the corresponding Bragg peak position does not depend on the field. Thus, the peak intensity of $(0, -1/2, 3/2)$ (AF) and $(0, -1/3, 1)$ (UUD) was measured versus field at fixed $\mathbf{Q}$. These Bragg peaks abruptly disappear when transitioning into the SDW (from AF) phase and D phase (from UUD) correspondingly. Therefore, the residual intensity measured at these $\mathbf{Q}$-points well above the phase transitions is purely due to background. A flat model of this background was fitted and removed from the data. The

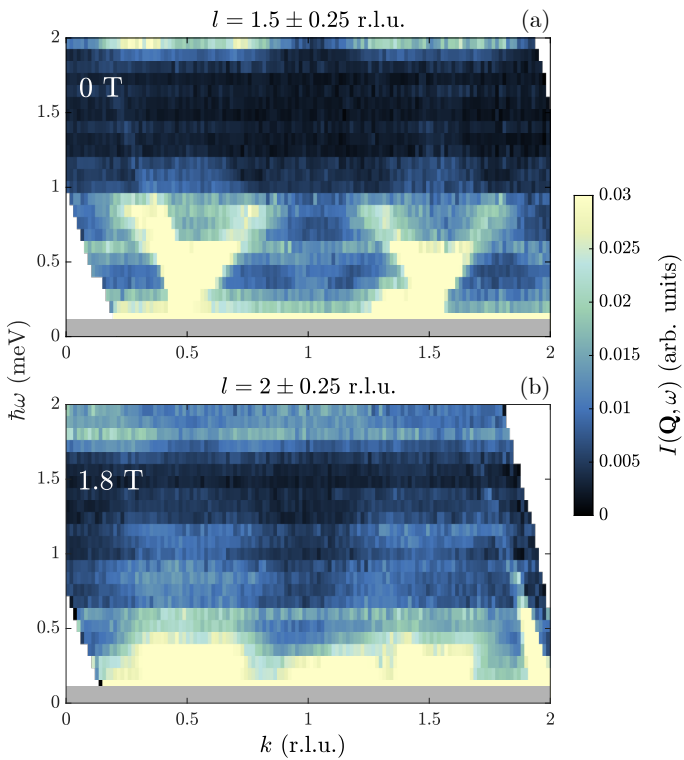

SUPP. FIG. 4. Raw datasets measured in the AF phase at 0 T (a) and in the SDW at 1.8 T (b). The integration ranges are shown in the figure. Gray areas mask regions of strong incoherent scattering.

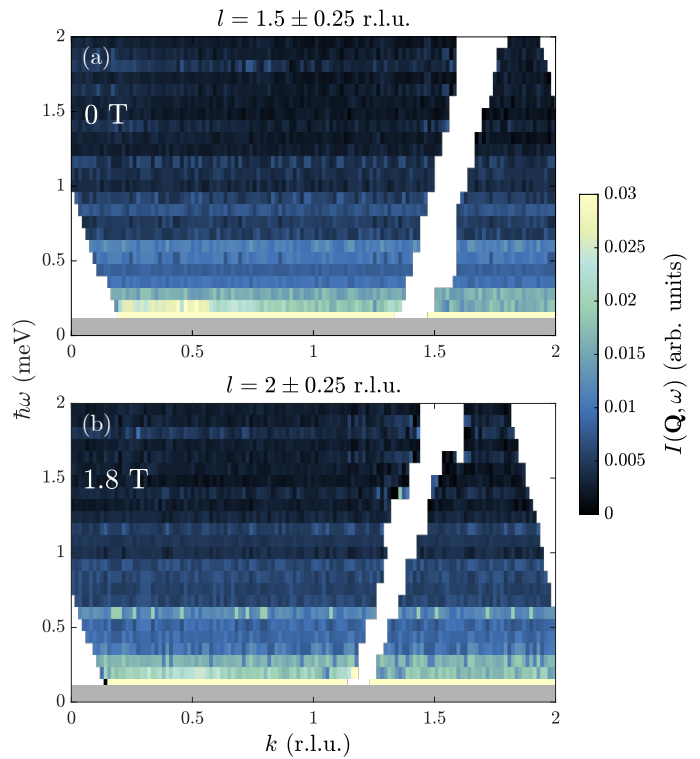

SUPP. FIG. 5. Background removed from Supp. Fig. 4, constructed as described in the text.

ordered moment is proportional to the square root of the Bragg peak amplitude, so the obtained peak intensities were normalized to the known ordered moment value from the magnetic structure refinement at a particular field value.

In the SDW phase, the Bragg peak position moves along the $k$-direction in changing fields. We thus performed broad $k$-scans at 8 values of field to track peak positions and amplitudes. A flat background was removed and the profile was fit to a gaussian function. $M_\mathbf{Q}$ is obtained from the square root of the peak amplitude, also normalized to the known magnetic structure.

### III. BACKGROUND SUBTRACTION PROCEDURE FOR CAMEA DATA

All the inelastic neutron scattering data presented in the main text are background subtracted (except for Fig. 4(e)). In this section the model adopted to describe the background is outlined.

Supp. Fig. 4 shows raw data at two different fields, integrated as in the main text. The first step in modeling the background was masking the portion of the data affected by a strong powder line, most likely originated from the magnet (clearly visible in Supp. Fig. 4 around $k = 1.4$ at 0 T and $k = 1.2$ at 1.8 T, and in Fig. 4(e) of the main text). Additional scattering emanating from Bragg peaks and dispersing in energy was observed. This was ascribed to Currat-Axe spurions, by simulating their expected trajectory in energy-momentum space with MJOLNIR [5] and comparing it with the observed one. Examples of spurious scattering can be seen in Supp. Fig. 4 emerging from $(0, 0.5, 1.5)$ (a) and dispersing up to 2 meV, or from $(0, 2, 2)$ in (b). In all the cases, Currat-Axe spurions were masked out.

For the remaining data, the background was modeled based on the physical assumption that no magnetic scattering is expected below the gap. Therefore the background dataset $(k, l, \hbar\omega)_{\text{bkgd}}$ is identical to the zero-field data for $\hbar\omega \leq 0.24$ meV (which is a conservative estimate based on the extracted gap value). Moreover, for $1.28 \leq \hbar\omega \leq 1.76$ meV, no relevant magnetic signal is observed, so this part of the zero-field dataset was included in $(k, l, \hbar\omega)_{\text{bkgd}}$ and replicated for $\hbar\omega > 1.76$ meV. For energies $0.24 < \hbar\omega < 1.28$ meV, the field dataset was used to extract the background. To this end, narrow cuts 0.1 r.l.u.$_k \times$ 0.1 r.l.u.$_l$ were taken at integer values of $k$ every 0.2 r.l.u. in $l$ were minimal or null magnetic signal is present. These were subsequently interpolated over the whole $kl$ space and the $(k, l, \hbar\omega)_{\text{bkgd}}$ dataset was integrated and point-to-point subtracted from each cut considered. As a title of example, the particular background subtracted from Supp. Fig. 4 to obtain Fig. 4 in the main text is shown in Supp. Fig. 5.

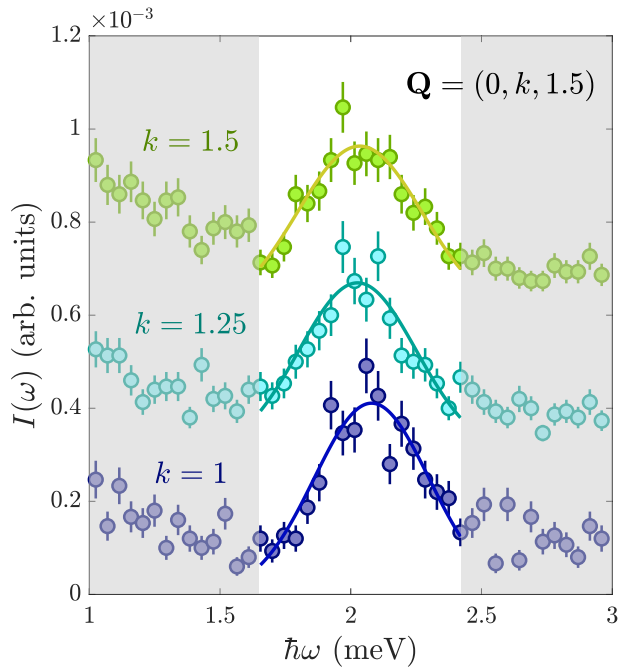

SUPP. FIG. 6. Intensity-energy profile of constant-**Q** scans for momentum transfer $\mathbf{Q} = (0, k, 1.5)$ at different $k$ measured on TASP. Solid lines are gaussian fits to the data, while gray areas hide data excluded from the fit. Light blue and green curves have been shifted upward for visibility by $3 \times 10^{-4}$ and $6 \times 10^{-4}$ arb. units respectively.

## IV. HIGHER ENERGY NON-DISPERSIVE LEVELS

From Fig. 2 of the main text an excess of intensity is visible at the upper edge of the spectrum, interpreted as transition between the Kramer doublets of $S = 3/2$ Cobalt split by the anisotropy. However, from the zero-field data, these levels are barely visible, with the tail of their intensity spilling at lower energies. In this section we provide additional data regarding these excitations.

Supp. Fig. 6 shows a few constant-**Q** scans performed on the triple-axis instrument TASP at PSI (Switzerland). The sample used for this experiment is a single crystal $m \sim 0.6$ g aligned in the $(0, k, l)$ scattering plane. All the measurements were performed at the base temperature of a dilution refrigerator and at zero-field. The particular data shown were measured at fixed $E_f = 4.66$ meV with curved analyzer, yielding and energy resolution $\sim 0.28$ meV FWHM at the elastic positions. The scans extend all the way up to 3 meV and confirm the presence of the intensity excess at higher energy observed on CAMEA. In particular, gaussian fits on these intensity profiles show no appreciable dispersion within the energy resolution of the experiment.

## V. ESTIMATING $J'$

### A. Linear spin-wave theory calculations (LSWT)

Linear spin-wave theory calculations were performed using the SpinW package [6]. These were performed in zero-field for all the magnetic atoms in the magnetic unit cell, based on the Hamiltonian (2) proposed in the main text. The magnetic structure was fixed to the experimentally determined one represented in Fig. 1 of the main text. By using $J = 0.8$ meV, $J' = 0.35$ meV, $\delta = 0.1$, $\Delta_{XY} = 0.75$ and $\delta' = 0.5$ the spectrum showed in Suppl. Fig. 7 was found. It essentially captures the presence of a spin gap $\sim 0.3$ meV and a substantially quenched transverse dispersion compared to an Heisenberg model with the same $J'$. It must be noticed that, however, the $l$-bandwith calculated from SpinW is substantially overestimated at $k = 0.75$ than at $k = 0.5$ r.l.u.. In the former case, indeed, the program interpolated between the upper and lower dispersion branch, with the former not being present in the actual data. The $l$-badnwidth at $k = 0.5$ is indeed flatter. In addition, increased spectral weight is expected at half-integer $k$, with vanishing intensity at $k$ integer.

Of course, since LSWT is a semiclassical approach, it is certainly ill-defined for systems involving strong quantum fluctuations like $Cs_2CoBr_4$. These results can be expected to be a good approximation to the spectrum only at the low energies, but not at the high ones. This is well known from multiple experimental examples including KCuF$_3$ and isostructural $Cs_2CuCl_4$ materials [7, 8]. Thus, we mostly use the gap value and the $k-$bandwidth as guidance for the approximate Hamiltonian parameters. Similar ideas are also adopted in the basic quantum model, discussed in the next section.

### B. Random Phase Approximation (RPA)

Another way of quantifying the bandwidth ratio in $k$- and $l$-directions is based upon using an Ising chain model. In this case, the elementary excitations are the kinks (domain walls), whose dispersion reads:

$$\epsilon(\mathbf{k}) = \sqrt{\mathcal{D}^2 + 2\mathcal{D}J(\mathbf{k})}. \quad \text{(S.1)}$$

Here the "gap" parameter $\mathcal{D}$ would be related to the degree of the anisotropy. Both in-chain and intra-chain exchanges are treated perturbatively (allowing the kink to propagate either by two sites in the chain direction, or onto the adjacent chain along the zig-zag bond). This RPA-type approach [9, 10] leads to:

$$J(\mathbf{k}) = 2J \cos 2k_x b \quad \text{(S.2)}$$
$$+ 2J' \cos(k_x b + k_y c/2) + 2J' \cos(k_x b - k_y c/2).$$

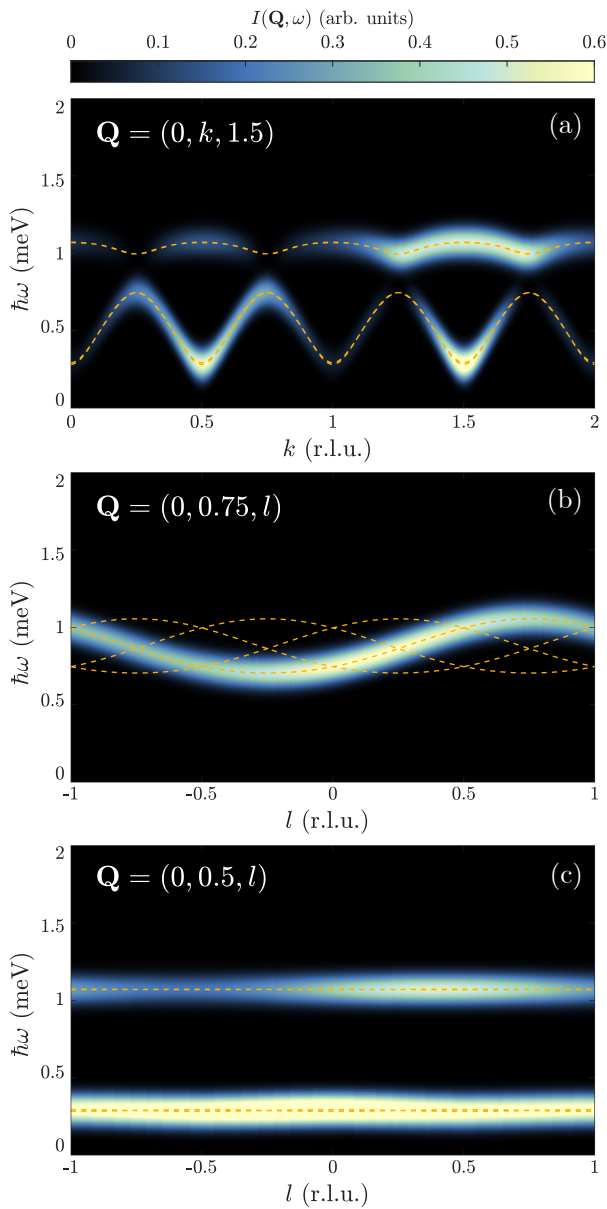

SUPP. FIG. 7. SpinW calculations for momentum transfer along $k$ and $l$ respectively, at two values of $k$ for the latter case. Orange dashed lines indicate the possible modes, with those carrying intensity shown in the colormap.

Here $x$ is the $b$, and $y$ is the $c$ direction of $Cs_2CoBr_4$. Since this is only a simple estimate, we neglect anisotropic exchange effects, apart from the presence of a gap in the spectrum. The bonds in Eqs. (S.1,S.2) are assumed Heisenberg-like.

In a neutron scattering process, only a pair of kinks is excited. Therefore, from the energy and momentum conservation laws, it follows

$$\hbar\omega(\mathbf{q}) = \epsilon(\mathbf{k}) + \epsilon(\mathbf{q}-\mathbf{k}). \quad (\text{S.3})$$

The upper and lower $\hbar\omega(\mathbf{q})$ boundaries of the two-particle continuum (S.3) are straightforwardly found, which allows the extraction of the bandwidths $\hbar\Delta\omega_k$ and $\hbar\Delta\omega_l$ at the lower edge. In our case, this was done numerically.

Specifically, we have performed the calculation for different $J'/J$ ratios in two different limits: $\mathcal{D} = 100J$, for which the gap is large and RPA-like treatment is well justified, and for $\mathcal{D} \simeq 6J$ limit in which the kink becomes nearly gapless and RPA is ill-defined. The results are shown in Supp. Fig. 8.

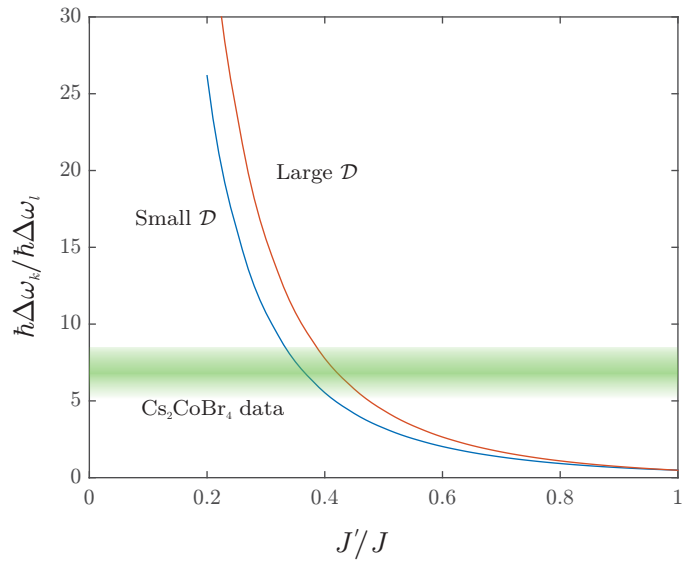

SUPP. FIG. 8. Bandwidth ratio in the RPA approximation, in the large and small gap limits respectively. The green area indicates the experimentally found bandwidth ratio.

In either case, the observed experimental bandwidth ratio $\simeq 7$ points to a ratio $J'/J \simeq 0.4$. Similar to the LSWT estimates, we would expect that allowing for the anisotropic natures of the bonds would suppress $\hbar\Delta\omega_l$ even more, and thus 0.4 is rather the lowest feasible value for the exchange ratio.

## VI. TOMONAGA-LUTTINGER LIQUID DYNAMICS AND THE SDW STATE

As seen from Supp. Fig. 9(top), magnetic excitations in a magnetized spin chain occur at incommensurate positions $Q^\pm = k_0 + 1/2 \pm m/2$ after each lattice peak $k_0$ (we are referring to the $k$-component of the wavevector here, assuming the "virtual" chain running along the $\mathbf{b}$ direction). These pairs of excitations are gapless and no Bragg peak is expected at $Q^\pm$. In actual materials like $BaCo_2V_2O_8$ with the ordered SDW phase these equidistant soft modes are accompanied by the magnetic Bragg peaks at the corresponding positions [11, 12].

On the other hand, the measured excitations in the SDW state of $Cs_2CoBr_4$ are similar to the situation depicted in Supp. Fig. 9(bottom). Due to crystal symme-

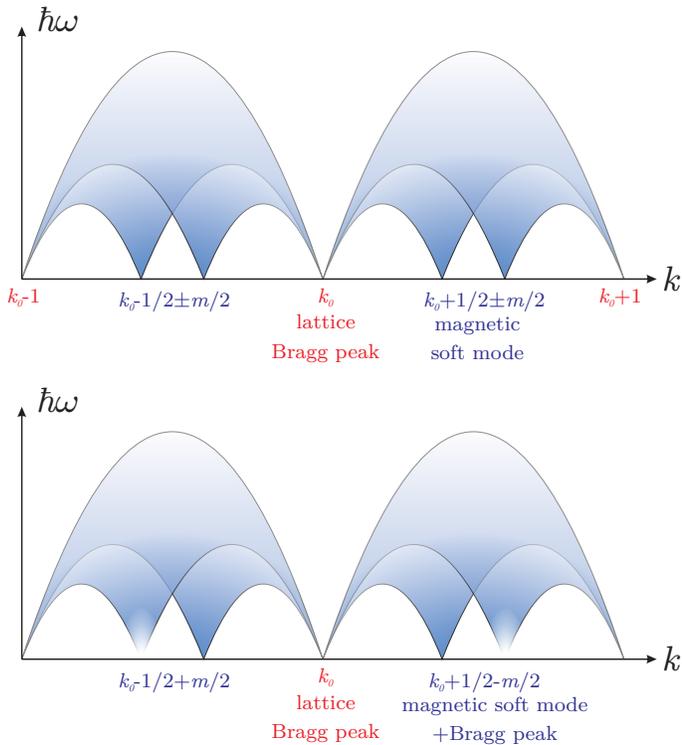

try, only every second lattice peak is allowed. In addition, a pair of excitations is not visible, with the surviving one associated to a magnetic Bragg peak.

SUPP. FIG. 9. Top: cartoon of $\mathcal{S}_{zz}$ magnetic excitations in the magnetized spin chain. Lattice peaks are denoted in red. Bottom: similar picture for Cs$_2$CoBr$_4$. Only every second $k_0$ lattice peak is possible, some soft modes are gone, and the remaining ones are accompanied by the magnetic Bragg peaks.



\end{document}